\documentclass[reprint,nofootinbib,showpacs]{revtex4}
\usepackage{graphicx}  
\usepackage{dcolumn}   
\usepackage{bm}        
\usepackage{amssymb}
\usepackage{hyperref}

\begin{document}

\title{\bf Predictions of $B_c \to (D,\,D^{\ast})\,l\,\nu$ decay observables in the standard model}
\author{Rupak~Dutta${}^{}$}
\email{rupak@phy.nits.ac.in}
\affiliation{
${}$National Institute of Technology Silchar, Silchar 788010, India\\
}

\begin{abstract}
We study $B_c \to (D,\,D^{\ast})\,l\,\nu$ semileptonic decays mediated via $b \to u\,l\,\nu$ charged current interactions using the 
$B_c \to (D,\,D^{\ast})$ transition form factors obtained in the covariant light front quark model. We give predictions 
on the branching ratios, ratio of branching ratios, and various asymmetries pertaining to these decay modes within the standard model.
These results can be tested in the ongoing or in the future experiments and can provide complimentary information regarding the observed
anomalies in $B \to (D,\,D^{\ast})\tau\nu$ and $B_c \to J/\Psi\tau\nu$ decays. 
\end{abstract}

\maketitle

\section{Introduction}
\label{int}
Flavor anomalies present in various $B$ meson decays compelled physicist to look for beyond the standard model~(SM) physics. The deviation
from the SM prediction observed in the measured ratio
of branching ratios $R_D$ and $R_{D^{\ast}}$~\cite{Lees:2013uzd,Huschle:2015rga,Sato:2016svk,Hirose:2016wfn,Aaij:2015yra} in 
$B \to (D,\,D^{\ast})\,l\nu$ decays and very recently measured $R_{J/\Psi}$~\cite{Aaij:2017tyk} in 
$B_c \to J/\Psi\,l\,\nu$ decays provide hints of lepton flavor universality violation. Currently, the discrepancy in $R_D$ and $R_{D^{\ast}}$
with the SM prediction~\cite{Lattice:2015rga,Na:2015kha,Aoki:2016frl,Bigi:2016mdz,Bernlochner:2017jka,Jaiswal:2017rve,Wang:2017jow,
Fajfer:2012vx,Bigi:2017jbd} stands at 
around $3.78\sigma$~\cite{Amhis:2016xyh}. Similarly, a deviation of around $1.3\sigma$ is observed in 
$R_{J/\Psi}$ from the SM prediction~\cite{Ivanov:2005fd,Wen-Fei:2013uea,Dutta:2017xmj,Cohen:2018dgz}. Furthermore, a deviation of around 
$1.5\sigma$~\cite{Ciezarek:2017yzh} is observed 
in $\mathcal B(B \to \tau\nu)$ decays mediated via $b \to u\,l\,\nu$ charged current interactions as well.
If this persists in future experiments, it would be a definite hint of existence of beyond the SM physics. 

Study of semileptonic $B_c$ meson decays will play an important role in the future. Being consist of two heavy quarks $b$ and $c$, the weak
decays of $B_c$ 
meson can occur via $b \to (u,\,d,\,c,\,s)$ and $c \to (u,\,d,\,s)$ transition decays. Although, measurements on non-leptonic decays of
$B_c$ meson exist~\cite{Patrignani:2016xqp,Aaij:2016xas}, however, for semileptonic decays, so far, LHCb has measured only the ratio
of branching ratios $R_{J/\Psi}$ 
in $B_c \to J/\Psi\,l\,\nu$ decays. With more and more data accumulated by the LHCb on $B_c$ meson, 
detection of semileptonic $B_c$ meson decays to various other mesons mediating via $b \to (c,\,u)\,l\,\nu$ charged current and 
$b \to s\,l\,\bar{l}$ neutral current interactions will be feasible. It should be
noted that study of various $B_c$ meson decays will act as complimentary decay channels to similar decays in the $B$ meson sector. 
Moreover, study of semileptonic $B_c$ meson decays both theoretically and experimentally are not only important but also necessary to 
maximize future sensitivity to new physics~(NP) contributions. Again, a precise determination of $B_c$ semileptonic branching ratio will be 
crucial to determine the not so well known CKM matrix elements as well. 

The semileptonic $B_c \to (D,\,D^{\ast})\,l\,\nu$ decay amplitude can be factorized into leptonic and hadronic matrix elements, where, the 
leptonic matrix element can be 
easily evaluated. The hadronic matrix element, on the other hand, depends on various meson to meson transition form factors that involve
QCD in the non perturbative regime. Several theoretical approaches exist such as the QCD sum rules, the covariant light front quark model,
the relativistic constituent quark model, and the non relativistic QCD to evaluate $B_c \to M$ transition form factors~\cite{Du:1988ws,
Colangelo:1992cx,Nobes:2000pm,Ivanov:2000aj,Kiselev:2000pp,Kiselev:2002vz,Ebert:2003cn,Ebert:2003wc}. We follow the covariant
light front quark model adopted in Ref.~\cite{Wang:2008xt} for the $B_c \to D$ and $B_c \to D^{\ast}$ transition form factors. Our main aim 
is to provide prediction on various observables related to $B_c \to (D,\,D^{\ast})\,l\,\nu$ decays in the SM. We give first prediction on 
various observables such as the ratio of branching ratios, lepton side forward backward asymmetry, polarization fraction of the charged 
lepton, convexity parameter, longitudinal and transverse polarization fraction of $D^{\ast}$ meson and the forward backward asymmetry of the 
transversely polarized $D^{\ast}$ meson.

The paper is organized as follows. In section~\ref{efh}, we start with the effective weak Hamiltonian for $b \to u\,l\,\nu$ transition decays
and subsequently write down the general form of the differential decay distribution for the $B_c \to (D,\,D^{\ast})\,l\,\nu$ semileptonic
decays. We write down explicit expressions of various observables pertaining to $B_c \to D\,l\,\nu$ and $B_c \to D^{\ast}\,l\,\nu$
decays in section~\ref{bcdlnu} and section~\ref{bcdslnu}, respectively. We present our results and discussion in section~\ref{rd} with
a brief summary and conclusion in section~\ref{con}.

\section{Formalism}
\subsection{Effective weak Hamiltonian}
\label{efh}
Effective field theory approach is the natural way to separate the long distance and the short distance effects coming from different scales
involved in the electroweak interaction~\cite{Buchalla:1995vs,Buras:1998raa}. Another advantage of using effective field theory is that it can 
be easily extended to include NP effects by introducing new operators~\cite{Cirigliano:2009wk,Bhattacharya:2011qm}.
At the quark level, the $B_c \to (D,\,D^{\ast})\,l\,\nu$ decays are governed by the effective weak Hamiltonian in which the heavy degrees of 
freedom are integrated out. The effects of the heavy degrees of freedom are absorbed into the short distance coefficients. The effective weak 
Hamiltonian for the $b \to u\,l\,\nu$ quark level transition, within the SM, can be written as
\begin{eqnarray}
\label{effh}
\mathcal H_{eff} = \frac{G_F}{\sqrt{2}}\,V_{ub}\,\bar{u}\,\gamma_{\mu}(1-\gamma_5)\,b\,\bar{l}\,\gamma^{\mu}\,(1-\gamma^5)\,\nu_l\,,
\end{eqnarray}
where $G_F$ is the Fermi coupling constant and $V_{ub}$ is the relevant Cabibbo-Kobayashi-Maskawa~(CKM) matrix element. Using the effective
Hamiltonian of Eq.~\ref{effh}, the three body differential decay distribution for the $B_c \to (D,\,D^{\ast})\,l\,\nu$ decays can be
expressed in terms of two kinematic variables $q^2$ and $\theta_l$ as follows:
\begin{eqnarray}
\frac{d\Gamma}{dq^2 d\cos\theta_l}= \frac{G_{F}^2 |V_{ub}|^2 |\vec{P}_{(D,D^{\ast})}|}{2^9\,\pi^3\,m_{B_c}^2} \left(1-\frac{m_{l}^2}{q^2}
\right)\,L_{\mu \nu} H^{\mu \nu}\,,
\end{eqnarray} 
where $q^2$ denotes lepton invariant mass squared and $\theta_l$ denotes the the angle between the $\vec{P}_{D,\,D^{\ast}}$ and lepton three 
momentum vector in the $l-\nu$ rest frame. Here 
$|\vec{P}_{(D,D^{\ast})}|= \sqrt{\lambda (m_{B_c}^2, m_{(D,D^{\ast})}^2, q^2)}/2m_{B_c}$ denotes the three momentum vector of the outgoing 
meson, where, $\lambda(a,b,c)=a^2+b^2+c^2-2(ab+bc+ca)$. The covariant contraction of $L_{\mu\nu}$~(lepton tensor) and 
$H^{\mu\nu}$~(hadron tensor) can be obtained by using the helicity techniques. A detail discussion of the helicity techniques can be found in 
Refs.~\cite{Korner:1989qb,Kadeer:2005aq}. 

\subsection{$B_c \to D\,l\,\nu$ decay observables}
\label{bcdlnu}
The full angular distribution of three body $B_c \to D\,l\,\nu$ decays can be written as 
\begin{eqnarray}
\label{dlnutheta}
\frac{d\Gamma^D}{dq^2\,d\cos\theta_l} = 2\,N\,|\vec{P}_{D}|\,\Bigg\{H_0^2\,\sin^2\theta_l\, + \frac{m_l^2}{q^2}\,
\Big[(H_0\,\cos\theta_l - H_{t})^2\Big]\Bigg\}\,,
\end{eqnarray}
with
\begin{eqnarray}
&&N = \frac{G_F^2\,|V_{u\, b}|^2\,q^2}{256\,\pi^3\,m_{B_c}^2}\,\Big(1 - \frac{m_l^2}{q^2}\Big)^2\,, \qquad\qquad
H_0 = \frac{2\,m_{B_c}\,|\vec{P}_{D}|}{\sqrt{q^2}}\,f_{+}(q^2)\,, \qquad\qquad
H_t = \frac{m_{B_c}^2 - m_{D}^2}{\sqrt{q^2}}\,f_0(q^2)\,, 
\end{eqnarray}
where $f_0(q^2)$ and $f_+(q^2)$ are the relevant $B_c \to D$ transition form factors.
By performing the integration over $\cos\theta_l$, we obtain the differential decay rate as follows:
\begin{eqnarray}
\frac{d\Gamma^D}{dq^2} = \frac{8\,N\,|\vec{P}_{D}|\,}{3}\Bigg\{\,H_0^2\,\Big(1 + \frac{m_l^2}{2\,q^2}\Big) + 
\frac{3\,m_l^2}{2\,q^2}\,H_t^2\Bigg\}\,.
\end{eqnarray}
We also investigate the $q^2$ dependent differential ratio defined by
\begin{eqnarray}
\label{rbcd}
&&R(q^2)=\frac{{\rm DBR}(B_c \to D\tau\nu)(q^2)}{{\rm DBR}(B_c \to D\,l\,\nu)(q^2)}\,,
\end{eqnarray}
where, $l$ is either an electron or muon and ${\rm DBR}(B_c \to D\tau\nu)(q^2) = \frac{d\Gamma^D}{dq^2}/\Gamma$. Note that 
$\Gamma = 1/\tau_{B_c}$ denotes the total decay width of the $B_c$ meson. Furthermore, we also explore several other observables such as 
the lepton side forward backward asymmetry $A_{FB}^l(q^2)$, the longitudinal polarization fraction of the charged lepton $P^l(q^2)$, 
and the convexity parameter $C_F^l(q^2)$ for the $B_c \to D\,l\,\nu$ decays.
The forward backward asymmetry in the angular distribution is defined as
\begin{eqnarray}
\label{afbd}
A^l_{FB}(q^2) = \frac{\Big(\int_{-1}^{0}-\int_{0}^{1}\Big)d\cos\theta\frac{d\Gamma^D}{dq^2\,d\cos\theta}}{\frac{d\Gamma^D}{dq^2}} \, =
\frac{3m_{l}^{2}}{2q^2}\frac{H_0\,H_t}{H_0^2\,\Big(1+\frac{m_l^2}{2\,q^2}\Big)+\frac{3\,m_l^2}{2q^2}\,H_t^2}\,.
\end{eqnarray}
Similarly, the longitudinal polarization fraction of the charged lepton is defined as
\begin{eqnarray}
\label{pd}
P^l(q^2)=\frac{d\Gamma^D(+)/dq^2 - d\Gamma^D(-)/dq^2}{d\Gamma^D/dq^2} = \frac{\frac{m_l^2}{2\,q^2}\,(H_0^2+3\,H_t^2) - H_0^2}
{\frac{m_l^2}{2\,q^2}\,(H_0^2+3\,H_t^2) + H_0^2}\,,
\end{eqnarray}
where $d\Gamma^D(+)/dq^2$ and $d\Gamma^D(-)/dq^2$ define the differential decay rate of the positive and negative helicity leptons, 
respectively.
Again, by taking second derivative of Eq.~\ref{dlnutheta} with respect to $\cos\theta_l$, we obtain the convexity parameter. That is
\begin{eqnarray}
\label{cfd}
C_F^l(q^2) = \frac{1}{\left(d\Gamma^D/dq^2\right)}\frac{d^2}{d(\cos\theta)^2}\left[\frac{d\Gamma^D}{dq^2\,d\cos\theta}\right]\, = 
\frac{3}{2}\frac{H_0^2\Big(\frac{m_l^2}{q^2}-1\Big)}
{\Big[H_0^2\,\Big(1+\frac{m_l^2}{2\,q^2}\Big)+\frac{3\,m_l^2}{2\,q^2}\,H_t^2\Big]}\,.
\end{eqnarray}
We also give predictions on the average values of the ratio of branching ratios $R_{B_cD}$, forward backward asymmetry $A_{FB}^l$, the
longitudinal polarization fraction $P^l$, and the convexity parameter $C_F^l$ for the $B_c \to D\,l\,\nu$ decay mode by separately
integrating the numerator and the denominator of Eq.~\ref{rbcd}, Eq.~\ref{afbd}, Eq.~\ref{pd}, and Eq.~\ref{cfd} over $q^2$.
\subsection{$B_c \to D^{\ast}\,l\,\nu$ decay observables}
\label{bcdslnu}
We write the three body $B_c \to D^{\ast}\,l\,\nu$ differential~($q^2$, $\cos\theta_l$) decay distribution as
\begin{eqnarray}
\frac{d\Gamma^{D^{\ast}}}{dq^2\,d\cos\theta_l}= N\,|\vec{P}_{D^{\ast}}|\Bigg\{2\mathcal{A}_0^2\,\sin^2\theta_l 
 + \Big[\Big(1+\cos^2\theta_l\Big) + \frac{m_l^2}{q^2}\sin^2\theta_l\Big]\Big[\mathcal{A}_{\|}^2 + \mathcal{A}_{\bot}^2\Big]  
 -4\,\mathcal{A}_{\|}\mathcal{A}_{\bot}\cos\theta_l + \frac{2\,m_l^2}{q^2}\Big[\mathcal{A}_0\cos\theta_l - \mathcal{A}_t\Big]^2\Bigg\}\,.
\end{eqnarray}
By performing the integration over $\cos\theta_l$, we obtain
\begin{eqnarray}
\frac{d\Gamma^{D^{\ast}}}{dq^2}=\frac{8\,N\,|\vec{P}_{D^{\ast}}|}{3}\Big[\Big(\mathcal{A}^2_0+ \mathcal{A}^2_{\|}+ 
\mathcal{A}^2_{\bot}\Big)\Big(1 + \frac{m_l^2}{2\,q^2}\Big)+\frac{3\,m_l^2}{2\,q^2}
\,\mathcal{A}^2_t\Big]\,
\end{eqnarray}
with
\begin{eqnarray}
&&\mathcal{A}_0=\frac{1}{2\,m_{D^{\ast}}\,\sqrt{q^2}}\Big[\Big(\,m_{B_c}^2-m_{D^{\ast}}^2-q^2\Big)(m_{B_c}+m_{D^{\ast}})A_1(q^2)\,-\,
\frac{4\,m_{B_c}^2|\vec{p}_{D^{\ast}}|^2}{m_{B_c}+m_{D^{\ast}}}A_2(q^2)\Big]\,,\nonumber\\
&&\mathcal{A}_{\|}=\frac{2\,(m_{B_c}+m_{D^{\ast}})\,A_1(q^2)}{\sqrt{2}}\,, \qquad
\mathcal{A}_{\bot}=-\frac{4\,m_{B_c}\,V(q^2)\,|\vec{P}_{D^{\ast}}|}{\sqrt{2}\,(m_{B_c}+m_{D^{\ast}})}\,, \qquad
\mathcal{A}_{t}=\frac{2\,m_{B_c}|\vec{P}_{D^{\ast}}|\,A_0(q^2)}{\sqrt{2}}\,,
\end{eqnarray}
where $V(q^2)$, $A_0(q^2)$, $A_1(q^2)$, and $A_2(q^2)$ are the relevant $B_c \to D^{\ast}$ transition form factors.

Similar to $B_c \to D\,l\,\nu$ decay observables, we investigate differential ratio $R(q^2)$, forward backward asymmetry parameter 
$A_{FB}^l(q^2)$, longitudinal polarization fraction of the charged lepton $P^l(q^2)$, and the convexity
parameter $C_F^l(q^2)$ for the $B_c \to D^{\ast}\,l\,\nu$ decays. The explicit expressions are
\begin{eqnarray}
\label{q2_obs_ds}
&&R(q^2) = \frac{{\rm DBR}(B_c \to D^{\ast}\tau\nu)(q^2)}{{\rm DBR}(B_c \to D^{\ast}\,l\,\nu)(q^2)}\,,\qquad\qquad
l\in (e,\,\mu)\,,\nonumber \\
&&A_{FB}^{l}(q^2) = \frac{3}{2}\frac{\mathcal{A}_{\|}\mathcal{A}_{\bot} + \frac{m_{l}^{2}}{q^{2}}\mathcal{A}_{0} \mathcal{A}_{t}}
 {\Big(\mathcal{A}_{0}^2 + \mathcal{A}_{\|}^2 + \mathcal{A}_{\bot}^2 \Big) \Big( 1+ \frac{m_{l}^{2}}{2q^{2}} \Big) + 
 \frac{3\,m_{l}^{2}}{2\,q^{2}}\mathcal{A}_{t}^2}\,, \nonumber \\
&&P^{l}= \frac{\Big(\mathcal{A}_{0}^2 + \mathcal{A}_{\|}^2 + \mathcal{A}_{\bot}^2 \Big)\Big(\frac{m_{l}^{2}}{2q^{2}}-1 \Big) - 
\frac{3\,m_{l}^{2}}{2q^{2}}\mathcal{A}_{t}^2}{\Big(\mathcal{A}_{0}^2 + \mathcal{A}_{\|}^2 + \mathcal{A}_{\bot}^2 \Big) 
\Big( 1+ \frac{m_{l}^{2}}{2q^{2}} \Big) + \frac{3\,m_{l}^{2}}{2q^{2}}\mathcal{A}_{t}^2}\,, \nonumber \\
&&C_F^l= \frac{3}{4}\frac{\Big(\frac{m_l^2}{q^2}-1 \Big) \Big(2\,\mathcal{A}_{0}^2 - \mathcal{A}_{\|}^2 - 
\mathcal{A}_{\bot}^2 \Big)}{\Big(\mathcal{A}_{0}^2 + \mathcal{A}_{\|}^2 + \mathcal{A}_{\bot}^2 \Big) \Big( 1+ \frac{m_l^2}{2\,q^2}
\Big) +  \frac{3\,m_l^2}{2\,q^2} \mathcal{A}_{t}^2}\,.
\end{eqnarray}
We also explore the forward backward asymmetry for the transversely polarized $D^{\ast}$ meson $A_{FB}^T(q^2)$ obtained by dropping the 
$\mathcal A_0$ and $\mathcal A_t$ from $A_{FB}^{l}(q^2)$ of Eq.~\ref{obs_ds}. Similarly, we define the longitudinal and transverse 
polarization fractions $F_L^{D^{\ast}}(q^2)$ and $F_T^{D^{\ast}}(q^2)$ of the $D^{\ast}$ meson. The explicit expressions are as follows:
\begin{eqnarray}
\label{q2_obs_ds1}
&&A_{FB}^T(q^2) = \frac{3}{2}\frac{\mathcal{A}_{\|}\mathcal{A}_{\bot}}
 {\Big(\mathcal{A}_{\|}^2 + \mathcal{A}_{\bot}^2 \Big) \Big( 1+ \frac{m_{l}^{2}}{2\,q^{2}}\Big)}\,,\nonumber \\
&&F_L^{D^{\ast}}(q^2) = \frac{\mathcal A_0^2\Big(1 + \frac{m_l^2}{2\,q^2}\Big)+\frac{3\,m_l^2}{2\,q^2}\,A_t^2}{\Big(\mathcal{A}^2_0+ 
\mathcal{A}^2_{\|}+ \mathcal{A}^2_{\bot}\Big)\Big(1 + \frac{m_l^2}{2\,q^2}\Big)+\frac{3\,m_l^2}{2\,q^2}\,\mathcal{A}^2_t}\,, \nonumber \\
&&F_T^{D^{\ast}}(q^2) = \frac{\Big(\mathcal A_{\|}^2 + \mathcal A_{\bot}^2\Big)\Big(1 + \frac{m_l^2}{2\,q^2}\Big)}{\Big(\mathcal{A}^2_0+ 
\mathcal{A}^2_{\|}+ \mathcal{A}^2_{\bot}\Big)\Big(1 + \frac{m_l^2}{2\,q^2}\Big)+\frac{3\,m_l^2}{2\,q^2}\,\mathcal{A}^2_t}\,,
\end{eqnarray}
where $F_L^{D^{\ast}} + F_T^{D^{\ast}} = 1$. Similar to $B_c \to D\,l\,\nu$ decays in section~\ref{bcdlnu}, we also give perdictions on
the average values of all the observables defined in Eq.~\ref{q2_obs_ds} and Eq.~\ref{q2_obs_ds1}.

\section{Results}
\label{rd}
\subsection{Inputs}
\label{in}
For our numerical estimates, we use $m_D = 1.86486\,{\rm GeV}$, $m_{D^{\ast}}=2.00698\,{\rm GeV}$, and $m_{B_c} = 6.2751\,{\rm GeV}$. 
Similarly for the lepton masses, we use $m_e=0.5109989461\times 10^{-3}\,{\rm GeV}$ and $m_{\tau}=1.77682\,{\rm GeV}$. The mass of $b$ quark 
at the renormalization scale $\mu = m_b$ is taken to be $m_b(m_b) = 4.18\,{\rm GeV}$. For the
lifetime of $B_c$ meson, we use $\tau_{B_c} = 0.507d-12\,{\rm S}$. We use Ref.~\cite{Patrignani:2016xqp} for all the above mentioned inputs.  
The CKM matrix element is taken to be $|V_{ub}| = (36.1\pm 3.2)\times 10^{-4}$~\cite{Patrignani:2016xqp}.
In order to make predictions on various observables, we need the information of 
various form factors related to $B_c \to D$ and $B_c \to D^{\ast}$ transitions. We follow Ref.~\cite{Wang:2008xt} and report in 
Table.~\ref{tab_ff} the relevant form factor inputs that are obtained in the covariant light-front quark model.
\begin{table}[htbp]
\centering
\begin{tabular}{|c|c|c|c|}
\hline
$F$ & $F(0)$ &$c_1$ &$c_2$ \\
\hline
\hline
$f_+^{B_c\,D}$ &$0.16\pm 0.03$ &$3.46\pm 0.31$ & $0.90\pm 0.08$ \\
\hline
$f_0^{B_c\,D}$ &$0.16\pm 0.03$ &$2.41\pm 0.28$ &$0.47\pm 0.06$ \\
\hline
$V^{B_c\,D^{\ast}}$ &$0.13\pm 0.03$ &$4.21\pm 0.39$ &$1.09\pm 0.10$ \\
\hline
$A_0^{B_c\,D^{\ast}}$ & $0.09\pm 0.01$ & $4.18\pm 0.40$ & $0.96\pm 0.11$ \\
\hline
$A_1^{B_c\,D^{\ast}}$ & $0.08\pm 0.01$ & $3.18\pm 0.37$ & $0.65\pm 0.08$ \\
\hline
$A_2^{B_c\,D^{\ast}}$ & $0.07\pm 0.01$ & $3.78\pm 0.35$ & $0.80\pm 0.07$ \\
\hline
\hline
\end{tabular}
\caption{$B_c \to D,\,D^{\ast}$ form factor inputs from Ref.~\cite{Wang:2008xt}.}
\label{tab_ff}
\end{table}
The momentum dependence of the form factors is parametrized as follows: 
\begin{eqnarray}
F(q^2) = F(0)\,\exp(c_1\,s + c_2\,s^2)\,,
\end{eqnarray}
where $s = q^2/m_{B_c}^2$ and $F$ represents the form factors $f_+$, $f_0$ and $V$, $A_0$, $A_1$, and $A_2$, respectively.

Uncertainties in the calculation of the $B_c \to (D,\,D^{\ast})\,l\,\nu$ decay amplitudes may come from two kinds of inputs.
First kind of inputs are very well known such as mass of mesons, mass of quarks, mass of leptons and lifetime of the $B_c$ meson. 
Second kind of inputs are not very well known hadronic parameters such as the CKM matrix element $|V_{ub}|$ and $B_c \to (D,\,D^{\ast})$ 
transition form factors. To estimate
the uncertainties associated with each observable, we perform a $\chi^2$ analysis based on the uncertainties coming from the not so
well known hadronic inputs only. Uncertainties associated with the well known inputs are omitted in our $\chi^2$ analysis. 
\subsection{SM prediction of $B_c \to D\,l\,\nu$ decay observables}
The SM prediction of $B_c \to D\,l\,\nu$ decay observables are reported in Table.~\ref{tab_d}. We use the central values of all the input
parameters to obtain the central values of each observables. To find the SM range, we define a $\chi^2$ as follows:
\begin{eqnarray}
\chi^2 = \sum_{i=1}^{7}\frac{(\mathcal O_i - \mathcal O_i^0)^2}{\sigma_{i}^2}\,,
\end{eqnarray}
where $\mathcal O_i = \Big(|V_{ub}|,\,f_0,\,f_+,\,c_1^{f_0},\,c_2^{f_0},\,c_1^{f_+},\,c_2^{f_+}\Big)$. Here $\mathcal O_i^0$ represents the 
central value of each parameter and $\sigma_i$ represents the $1\sigma$ uncertainty associated with each parameter. The range of each 
observable is obtained by demanding $\chi^2 \le 0.989$.
The branching ratio of $B_c \to D\,l\,\nu$ is found to be of order $10^{-4}$ for both $e$ and the $\tau$ modes which is in good agreement with
the values obtained in Ref.~\cite{Wang:2008xt}. The slight difference may come from different choices of the input parameters. The value of 
each observables for the $e$ mode differs significantly from that of the $\tau$ mode. As expected, we obtain $A_{FB}^e=0$, $P^e = -1$, and
$C_F^e = -1.5$ in the SM.
\begin{table}[htbp]
\centering
\begin{tabular}{|c|c|c||c|c|c|}
\hline
Observables & Central value &Range &Observables &Central value &Range \\
\hline
\hline
$\mathcal B(B_c \to D\,e\nu)\times 10^4$ &$0.277$ &$[0.183, 0.406]$ &$\mathcal B(B_c \to D\tau\nu)\times 10^4$ &$0.195$ &$[0.138, 0.264]$\\
\hline
$P^e$ &$-1.000$ &$-1.000$&$P^{\tau}$ &$-0.047$ &$[-0.263, 0.183]$\\
\hline
$A^e_{FB}$ &$0.000$ &$0.000$&$A^{\tau}_{FB}$ &$0.294$ &$[0.269, 0.301]$\\
\hline
$C^e_F$ &$-1.500$ &$-1.500$&$C^{\tau}_F$ &$-0.556$ &$[-0.433, -0.672]$\\
\hline
$R_{B_c\,D}$ &$0.703$ &$[0.589, 0.891]$ & & &\\
\hline
\hline
\end{tabular}
\caption{SM prediction of various observables for the $B_c \to D\,l\,\nu$ semileptonic decays.}
\label{tab_d}
\end{table}

\begin{figure}[htbp]
\begin{center}
\includegraphics[width=5.8cm,height=5.0cm]{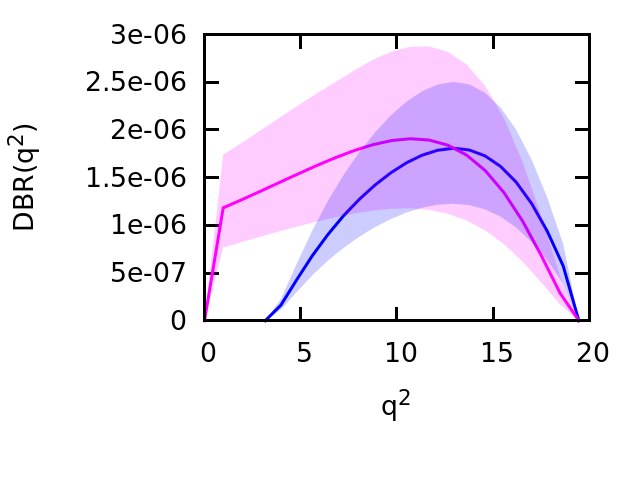}
\includegraphics[width=5.8cm,height=5.0cm]{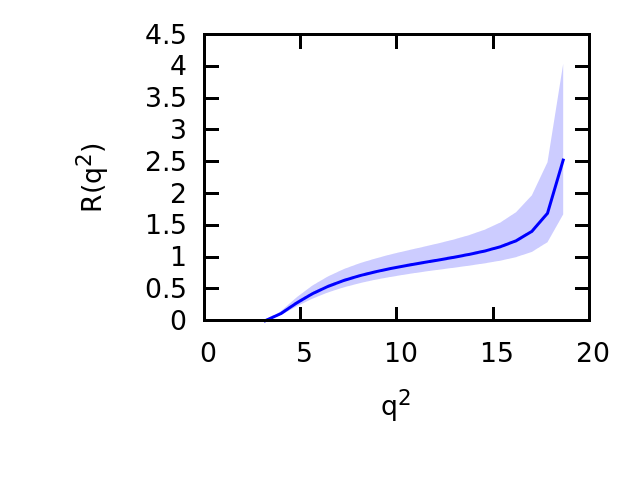}
\end{center}
\caption{Differential branching ratio ${\rm DBR}(q^2)$ and differential ratio $R(q^2)$ for $B_c \to D\,l\,\nu$ decays in the SM. The pink
band~($e$ mode) and the blue band~($\tau$ mode)
correspond to uncertainties in $|V_{ub}|$ and $B_c \to D$ form factor inputs. The pink and blue solid lines represent the SM prediction 
for the $e$ and the $\tau$ modes obtained using the central values of all the input parameters.}
\label{dbr_r_d}
\end{figure}

We show in Fig.~\ref{dbr_r_d} the differential branching ratio ${\rm DBR}(q^2)$ and differential ratio $R(q^2)$ as a function of $q^2$. 
The solid lines correspond to the central values of the input parameters, whereas, the bands correspond to
uncertainties in the CKM matrix element $|V_{ub}|$ and various $B_c \to D$ transition form factor inputs. It is observed that the differential
decay distribution is zero at $q^2 = m_l^2$ and $q^2 = q^2_{\rm max} = (m_{B_c} - m_D)^2$. This is obvious because at $q^2 = m_l^2$, $N$ goes
to zero and at $q^2 = (m_{B_c} - m_D)^2$, $\vec{P}_D$ goes to zero. The peak of the differential decay distribution for the $e$ 
mode~($3.0\times 10^{-6}\,{\rm GeV}^{-2}$) and the $\tau$ mode~($2.5\times 10^{-6}\,{\rm GeV}^{-2}$) occurs at around $q^2 = 11\,{\rm GeV}^2$ 
and $q^2 = 12.5\,{\rm GeV}^2$, respectively. For the differential ratio $R(q^2)$, we obtain the maximum value of $R(q^2) = 4$ at 
$q^2 = q^2_{\rm max}$. It is worth mentioning that the uncertainty is
less in $R(q^2)$ than ${\rm DBR}(q^2)$ due to cancellation of various uncertainties that are common to both the numerator and the denominator.

\begin{figure}[htbp]
\begin{center}
\includegraphics[width=5.8cm,height=5.0cm]{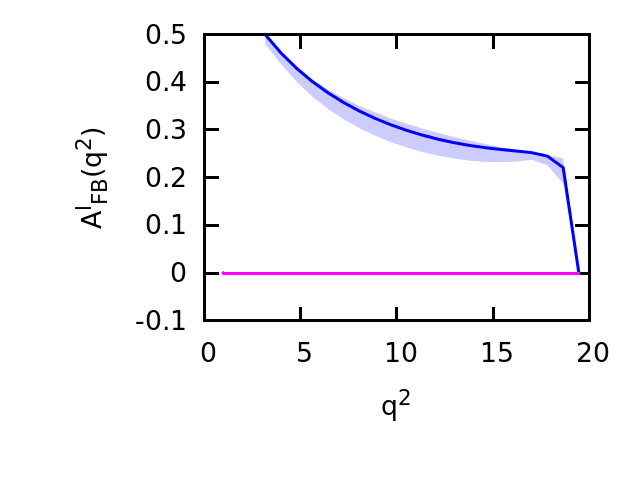}
\includegraphics[width=5.8cm,height=5.0cm]{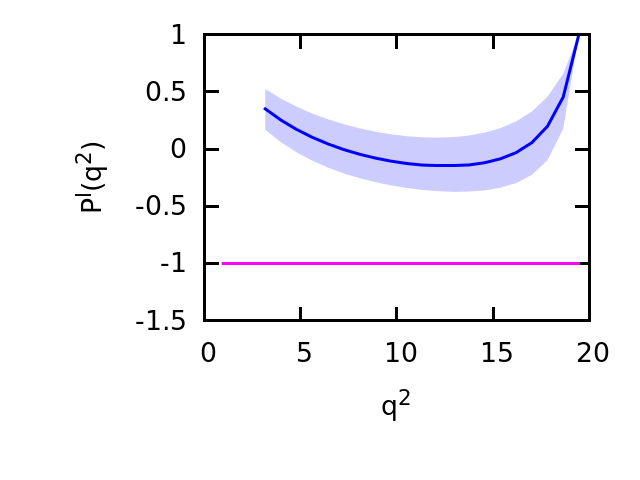}
\includegraphics[width=5.8cm,height=5.0cm]{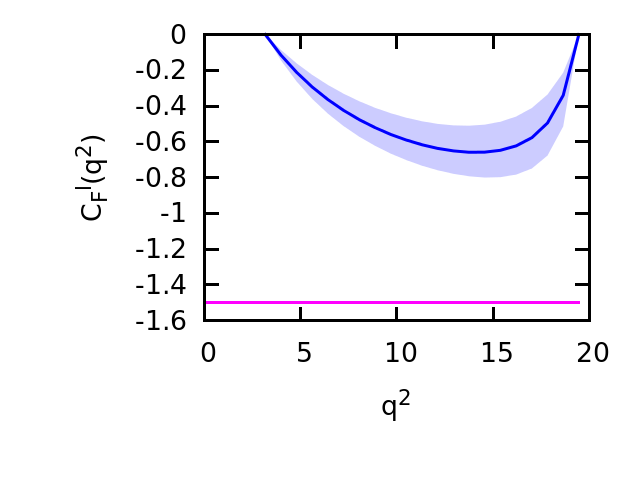}
\end{center}
\caption{Forward backward asymmetry $A_{FB}^l(q^2)$, longitudinal polarization fraction $P^l(q^2)$ and convexity parameter $C_F^l(q^2)$ for 
the $B_c \to D\,l\,\nu$ decays within the SM. Notations are same as in Fig.~\ref{dbr_r_d}.}
\label{obs_d}
\end{figure}

The $q^2$ dependence of the forward backward asymmetry $A_{FB}^l(q^2)$, longitudinal polarization fraction $P^l(q^2)$ and convexity 
parameter $C_F^l(q^2)$ in the SM is shown in Fig.~\ref{obs_d}. We observe that $A_{FB}^l(q^2)$, $P^l(q^2)$, and $C_F^l(q^2)$ for the $e$
mode remain constant throughout the whole $q^2$ region. However, for the $\tau$ mode, $A_{FB}^{\tau}(q^2)$ peaks~($0.5$) at low $q^2$ and 
gradually decreases as $q^2$ increases. It becomes zero at $q^2 = q^2_{\rm max}$. For the longitudinal polarization fraction of the $\tau$
lepton, $P^{\tau}(q^2)$ is around $0.5$ at low $q^2$, reaches its minimum value~($-0.47$) at $q^2 = 12.5\,{\rm GeV}^2$ and gradually increases
to $1$ at high $q^2$. It can assume 
both positive and negative values depending on the value of $q^2$. Considering the central blue solid line, we find two zero crossings in the 
$P^{\tau}(q^2)$ observable at $q^2 = 7.2\,{\rm GeV}^2$ and $q^2 = 16.5\,{\rm GeV}^2$, respectively. The convexity parameter $C_F^{\tau}(q^2)$
is zero at low and high $q^2$. It is found to be minimum~($-0.8$) at $q^2 = 15\,{\rm GeV}^2$. Although the forward backward asymmetry
$A_{FB}^{\tau}(q^2)$ is always positive, the convexity parameter $C_F^{\tau}(q^2)$ is always negative throughout the whole $q^2$ region.
\subsection{SM prediction of $B_c \to D^{\ast}\,l\,\nu$ decay observables}
We first report in Table.~\ref{tab_ds} the central value and the allowed range of each observables for the $B_c \to D^{\ast}\,l\,\nu$ 
semileptonic decays, where, $l$ is either an electron or a tau lepton, respectively. The central value of each observable is obtained
by using the central value of each input parameters, whereas, the allowed range is obtained by effecting a $\chi^2$ defined as
\begin{eqnarray}
\chi^2 = \sum_{i=1}^{13}\frac{(\mathcal O_i - \mathcal O_i^0)^2}{\sigma_{i}^2}\,,
\end{eqnarray}
where $\mathcal O_i$ denotes the CKM matrix element $|V_{ub}|$ and the $B \to D^{\ast}$ form factor inputs such as $V$, $A_0$, $A_1$, $A_2$
and the corresponding values of $c_1$ and $c_2$, respectively. Here $\mathcal O_i^0$ represents the
central value of each parameter and $\sigma_i$ represents the $1\sigma$ uncertainty associated with each parameter. We obtain the SM range
by demanding $\chi^2 \le 3.565$. The branching ratio $\mathcal B(B_c \to D^{\ast}\,l\,\nu)$ is of the order of $10^{-4}$ for both $e$ and
the $\tau$ modes, respectively. This is in good agreement with the results obtained in Ref.~\cite{Wang:2008xt}. In the SM, we obtain the
ratio $F_L^{D^{\ast}}/F_T^{D^{\ast}}$ to be around $2/3$ for the $e$ as well as the $\tau$ mode.
\begin{table}[htbp]
\centering
\begin{tabular}{|c|c|c||c|c|c|}
\hline
Observables & Central value &Range &Observables &Central value &Range \\
\hline
\hline
$\mathcal B(B_c \to D^{\ast}\,e\nu)\times 10^4$ &$0.395$ &$[0.217,0.715]$ & $\mathcal B(B_c \to D^{\ast}\tau\nu)\times 10^4$ &$0.238$ &
$[0.135,0.424]$\\
\hline
$P^e$ &$-1.000$ &$-1.000$&$P^{\tau}$ &$-0.654$ & $[-0.544,-0.715]$\\
\hline
$A^e_{FB}$ &$-0.365$ &$[-0.234, -0.504]$&$A^{\tau}_{FB}$ &$-0.222$ &$[-0.113,-0.345]$\\
\hline
$C^e_F$ &$-0.162$ &$[0.079, -0.319]$&$C^{\tau}_F$ &$-0.039$ &$[0.088,-0.124]$\\
\hline
$(A_{FB}^T)^e$ & $-0.613$ & $[-0.435,-0.720]$ & $(A_{FB}^T)^{\tau}$ & $-0.527$ & $[-0.364, -0.633]$\\
\hline
$(F_L^{D^{\ast}})^e$ & $0.405$ & $[0.298,0.475]$ &$(F_L^{D^{\ast}})^{\tau}$ & $0.412$ & $[0.334,0.465]$\\
\hline
$(F_T^{D^{\ast}})^e$ & $0.595$ & $[0.525,0.702]$ &$(F_T^{D^{\ast}})^{\tau}$ & $0.588$ & $[0.535,0.666]$\\
\hline
$R_{B_c\,D^{\ast}}$ &$0.603$ &$[0.573, 0.651]$ & & &\\
\hline
\hline
\end{tabular}
\caption{SM prediction of various observables for the $B_c \to D^{\ast}\,l\,\nu$ semileptonic decays.}
\label{tab_ds}
\end{table}

\begin{figure}[htbp]
\begin{center}
\includegraphics[width=5.8cm,height=5.0cm]{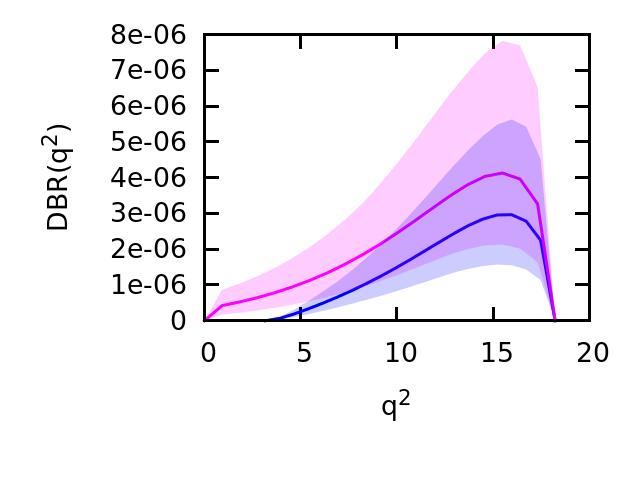}
\includegraphics[width=5.8cm,height=5.0cm]{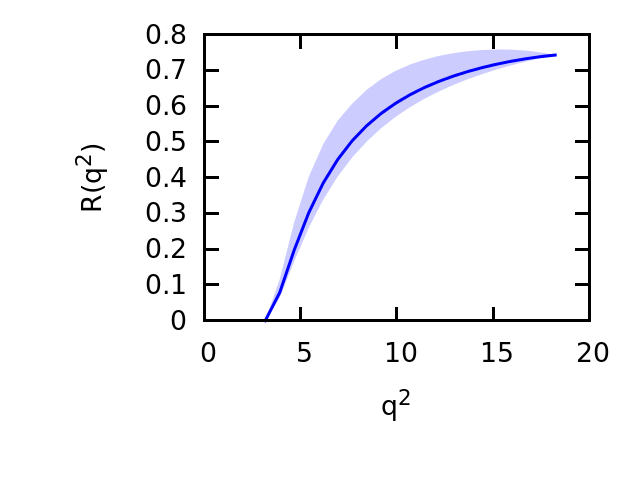}
\end{center}
\caption{Differential branching ratio ${\rm DBR}(q^2)$ and differential ratio $R(q^2)$ for $B_c \to D^{\ast}\,l\,\nu$ decays in the SM.
The pink band~($e$ mode) and the blue band~($\tau$ mode) correspond to uncertainties in $|V_{ub}|$ and $B_c \to D^{\ast}$ form factor inputs. 
The pink and blue solid lines represent 
the SM prediction for the $e$ and the $\tau$ modes obtained using the central values of all the input parameters.}
\label{dbr_r_ds}
\end{figure}

In Fig.~\ref{dbr_r_ds}, we show the differential branching ratio ${\rm DBR}(q^2)$ and differential ratio $R(q^2)$ as a function of the
lepton invariant mass squared $q^2$. The solid lines correspond to the central values of the input parameters, whereas, the bands correspond
to the uncertainties associated with the CKM matrix element $|V_{ub}|$ and various $B_c \to D^{\ast}$ transition form factor inputs.
The pink band represents the $e$ mode and the blue band represents the $\tau$ mode, respectively.
We observe that the ${\rm DBR}(q^2)$ for $B_c \to D^{\ast}\,e\,\nu$ peaks~($8\times 10^{-6}\,{\rm GeV}^{-2}$) at 
$q^2 = 15\,{\rm GeV}^2$, whereas, for the $B_c \to D^{\ast}\,\tau\,\nu$, it peaks~($5.5\times 10^{-6}\,{\rm GeV}^{-2}$) at 
$q^2 = 15\,{\rm GeV}^2$. We observe that as $q^2$ increases, $R(q^2)$ gradually increases and reach its maximum value $0.75$ at
$q^2 = q^2_{\rm max}$. It is evident that the uncertainty in $R(q^2)$ is greatly reduced in comparison to the uncertainty in 
${\rm DBR}(q^2)$. 

\begin{figure}[htbp]
\begin{center}
\includegraphics[width=5.8cm,height=5.0cm]{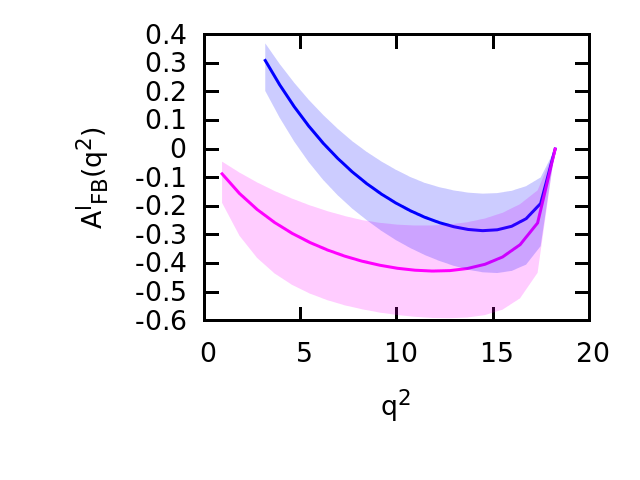}
\includegraphics[width=5.8cm,height=5.0cm]{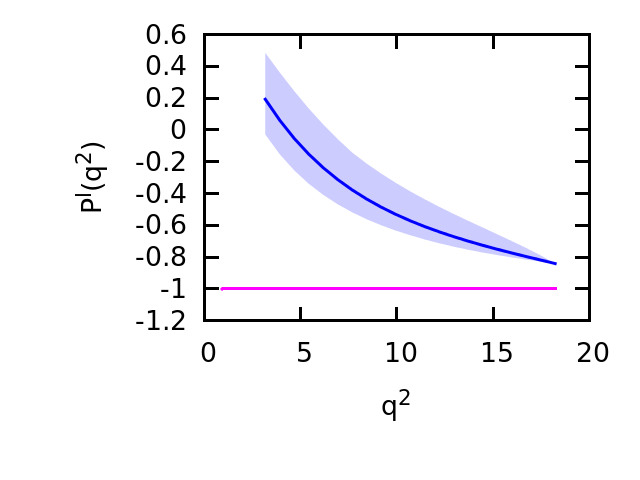}
\includegraphics[width=5.8cm,height=5.0cm]{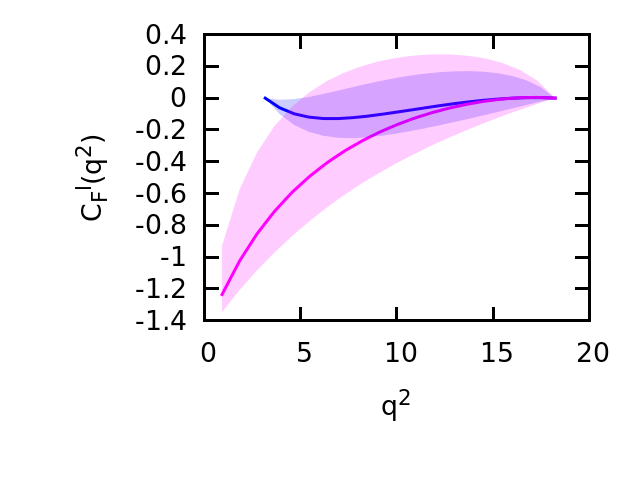}
\includegraphics[width=5.8cm,height=5.0cm]{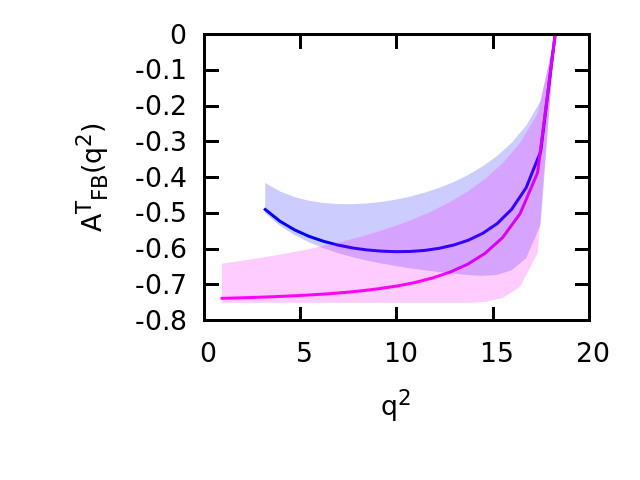}
\includegraphics[width=5.8cm,height=5.0cm]{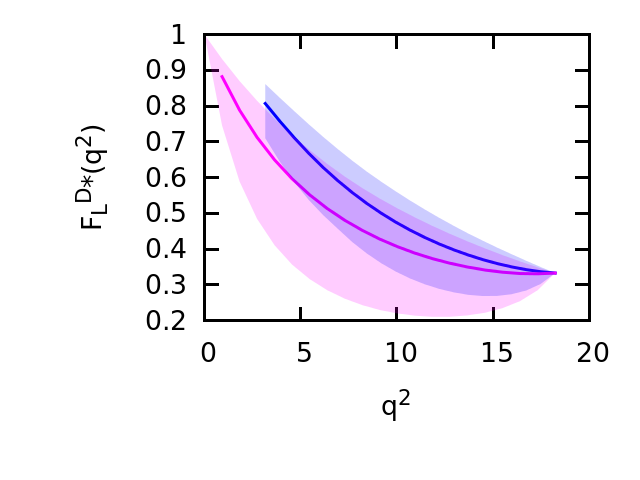}
\includegraphics[width=5.8cm,height=5.0cm]{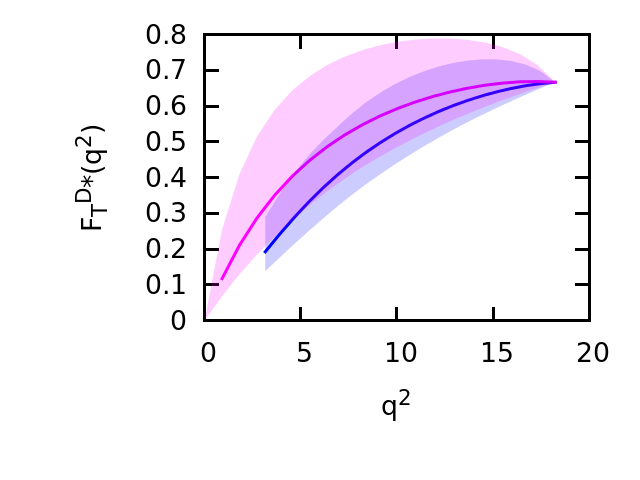}
\end{center}
\caption{Forward backward asymmetry $A_{FB}^l(q^2)$, longitudinal polarization fraction $P^l(q^2)$, convexity parameter $C_F^l(q^2)$, forward
backward asymmetry for the transversely polarized $D^{\ast}$ meson $A_{FB}^T(q^2)$, longitudinal polarization fraction of $D^{\ast}$ meson
$F_L(q^2)$ and the transverse polarization of the $D^{\ast}$ meson $F_T(q^2)$ for the
$B_c \to D^{\ast}\,l\,\nu$ decays within the SM. Notations are same as in Fig.~\ref{dbr_r_ds}.}
\label{obs_ds}
\end{figure}

We show in Fig.~\ref{obs_ds} various other $q^2$ dependent observables for the $B_c \to D^{\ast}\,l\,\nu$ decays. If we consider the central
values denoted by the solid lines, we find that there exists a zero crossing in the forward backward asymmetry parameter $A_{FB}^{\tau}(q^2)$ 
at $q^2 = 6.5\,{\rm GeV}^2$ above which it becomes negative. $A_{FB}^{\tau}(q^2)$ peaks~($0.3$) at low $q^2$ and reach its minimum of around 
$-0.4$ at $q^2 = 15\,{\rm GeV}^2$. It becomes zero at high $q^2$. However, $A_{FB}^{e}(q^2)$ remains negative throughout the whole $q^2$ 
region. Similarly, we observe a zero crossing in the $\tau$ polarization fraction
$P^{\tau}(q^2)$ at $q^2 = 4.3\,{\rm GeV}^2$. $P^{\tau}(q^2)$ is around $0.4$ at low $q^2$ and gradually decreases to around $-0.8$ at high
$q^2$. We observe that $P^{e}(q^2)$ remains constant~($-1.0$) in the whole $q^2$ region. It is observed that, if we consider the central 
solid lines, $C_F^l(q^2)$ for both $e$ and $\tau$ modes remains negative throughout the whole $q^2$ region. However considering the bands,
they can assume positive values depending on the value of $q^2$. The magnitude of $C_F^{e}(q^2)$ is much larger than the magnitude of 
$C_F^{\tau}(q^2)$. Magnitude of $C_F^{e}(q^2)$ can be as large as $-1.2$ at low $q^2$, whereas, $C_F^{\tau}(q^2)$ remains very small in the 
whole $q^2$ region. The asymmetry 
$A_{FB}^T(q^2)$ for both $e$ and $\tau$ mode are always negative. For the $e$ mode, $A_{FB}^T(q^2)$ can be as large as $-0.75$ below 
$q^2 = 15\,{\rm GeV}^2$, whereas, for the $\tau$ mode, it can assume large value of around $-0.65$ at $q^2 = 15\,{\rm GeV}^2$. In the SM, 
the longitudinal polarization
fraction of the $D^{\ast}$ meson $F_L^{D^{\ast}}(q^2)$ for the $\tau$ mode can be as large as $0.85$ at low $q^2$ and gradually decreases to 
about $0.3$ at high $q^2$. Similarly for the $e$ mode, it peaks~($1.0$) at low $q^2$ and decreases to about $0.2$ at high $q^2$. Again,
as expected, $F_T^{D^{\ast}}(q^2)$ for the $\tau$ mode is minimum~($0.15$) at low $q^2$ and it peaks~($0.7$) at high $q^2$. For the $e$ mode, 
$F_T^{D^{\ast}}(q^2)$ is zero at low $q^2$ and peaks~($0.8$) at high $q^2$. $F_L^{D^{\ast}}(q^2)$ and $F_T^{D^{\ast}}(q^2)$ remain positive
in the entire $q^2$ region.

\section{Conclusion}
\label{con}
We have investigated semileptonic decays of $B_c$ meson to $D$ and $D^{\ast}$ meson with a charged lepton and a neutrino in the final state.
We use the effective Hamiltonian for $b \to u\,l\,\nu$ quark level transition decays and give predictions on various observables for
these decay modes within the SM. We give first prediction of various observables such as the ratio of branching ratios, lepton side forward
backward asymmetry, longitudinal polarization fraction of the charged lepton, convexity parameter, forward backward asymmetry of the 
transversely polarized $D^{\ast}$ meson, longitudinal and transverse polarization fraction of the $D^{\ast}$ meson for these decays modes.
These results can be tested at future experiments and, in principle, can provide complimentary information regarding lepton flavor 
universality violation observed in various $B$ meson decays.

\end{document}